\documentclass[aps,nofootinbib,showpacs,twocolumn,floatfix]{revtex4}
\usepackage{epsf,amsfonts,amssymb,amsbsy}
\begin{document}
\title{Exclusive Decays and Lifetime of $\boldsymbol B_{\boldsymbol c}$ Meson
in QCD sum rules}
\author{V.V.Kiselev}
\email{kiselev@th1.ihep.su}
\affiliation{Russian State Research Center "Institute for High Energy
Physics",
Protvino, Moscow Region, 142280, Russia\\ Fax: +7-0967-744739}
\pacs{13.20.Gd,13.25.Gv,11.55.Hx}
\begin{abstract}
We summarize theoretical calculations of $B_c$ decays and lifetime in the
framework of QCD sum rules and compare the results with the estimates by the
methods of Operator Product Expansion in the inverse heavy quark masses as well
as of potential quark models. The agreement of estimates in the various
approaches is discussed. The features of $B_c$ decay modes are considered.
\end{abstract}
\maketitle

\hbadness=1500
\section{Introduction}

Decays of long-lived heavy meson $B_c$, containing two heavy quarks of
different flavors, were considered in the pioneering paper written by Bjorken
in 1986 \cite{Bj}. The report was devoted to the common view onto the decays of
hadrons with heavy quarks: the mesons and baryons with a single heavy quark,
the $B_c$ meson, the baryons with two and three heavy quarks. A lot of efforts
was recently directed to study the long-lived doubly heavy
hadrons\footnote{Reviews on the physics of $B_c$ meson and doubly heavy baryons
can be found in refs. \cite{revbc,revqq}, respectively.} on the basis of modern
understanding of QCD dynamics in the weak decays of heavy flavors in the
framework of today approaches\footnote{See the program on the heavy flavour
physics at Tevatron in \cite{BTeV}.}: the Operator Product Expansion, sum rules
of QCD and NRQCD \cite{NRQCD}, and potential models adjusted due to the data on
the description of hadrons with a single heavy quark. Surprisingly, the
Bjorken's estimates of total widths and various branching fractions are close
to what is evaluated in a more strict manner. At present we are tending to
study some subtle effects caused by the influence of strong forces onto the
weak decays of heavy quarks, which determines our understanding a probable fine
extraction of {\sf CP}-violation in the heavy quark sector.

Various hadronic matrix elements enter in the description of weak decays. So,
measuring the lifetimes and branching ratios implies the investigation of quark
confinement by the strong interactions, which is important in the evaluation of
pure quark characteristics: masses and mixing angles in the CKM matrix, all of
which enter as constraints on the physics beyond the Standard Model. More
collection of hadrons with heavy quarks provides more accuracy and confidence
in the understanding of QCD dynamics and isolation of bare quark values.

So, a new lab for such investigations is a doubly heavy long-lived quarkonium
$B_c$ recently observed for the first time by the CDF Collaboration \cite{cdf}.
The measured $B_c$ lifetime is equal to
$$
\tau[B_c] = 0.46^{+0.18}_{-0.16}\pm 0.03\; {\rm ps,}
$$
which is close to the value expected by Bjorken.

The $B_c$ meson allows us to use such advantages like the nonrelativistic
motion of $\bar b$ and $c$ quarks similar to what is know in the heavy
quarkonia $\bar b b$ and $\bar c c$, and suppression of light degrees of
freedom: the quark-gluon sea is small in the heavy quarkonia. These two
physical conditions imply two small expansion parameters for $B_c$:
\begin{itemize}
\item
the relative velocity of quarks $v$,
\item
the ratio of confinement scale to the heavy quark mass $\Lambda_{QCD}/m_Q$.
\end{itemize}
The double expansion in $v$ and $1/m_Q$ generalizes the HQET approach
\cite{HQET} to what is called as NRQCD \cite{NRQCD}. Moreover, the energy
release in the heavy quark decays determines the $1/m_Q$ parameter to be the
appropriate quantity for the Operator Product Expansion (OPE) and justifies the
use of potential models (PM) in the calculations of hadronic matrix elements,
too. The same arguments ensure the applicability of sum rules (SR) of QCD and
NRQCD.

The $B_c$ decays were, at first, calculated in the PM
\cite{PMBc,PMK,PML,PML2,vary,chch,ivanov,ISGW2,narod,CdF}, wherein the
variation of techniques results in close estimates after the adjustment on the
semileptonic decays of $B$ mesons. The OPE evaluation of inclusive decays gave
the lifetime and widths \cite{OPEBc}, which agree with PM, if one sums up the
dominating exclusive modes. That was quite unexpected, when the SR of QCD
resulted in the semileptonic $B_c$ widths \cite{QCDSRBc}, which are one order
of magnitude less than those of PM and OPE. The reason was the valuable role of
Coulomb corrections, that implies the summation of $\alpha_s/v$ corrections
significant in the heavy quarkonia, i.e. in the $B_c$ \cite{PML,KT,KLO,KKL}. At
present, all of mentioned approaches give the close results for the lifetime
and decay modes of $B_c$ at similar sets of parameters. Nevertheless, various
dynamical questions remain open:
\begin{itemize}
\item
What is the appropriate normalization point of non-leptonic weak lagrangian in
the $B_c$ decays, which basically determines its lifetime?
\item
What are the values of masses for the charmed and beauty quarks?
\item
What are the implications of NRQCD symmetries for the form factors of $B_c$
decays and mode widths?
\item
How consistent is our understanding of hadronic matrix elements, characterizing
the $B_c$ decays, with the data on the other heavy hadrons?
\end{itemize}
In the present work we shortly review the $B_c$ decays by summarizing the
theoretical predictions including new calculations on the exclusive decays in
the framework of QCD sum rules and discuss how direct experimental measurements
can answer the questions above.

In Section \ref{2} we recollect the inclusive estimates of $B_c$ lifetime in
various techniques: the Operator Product Expansion combined with the effective
theory of heavy quarks, the Potential Model approach and the QCD sum rules.
Section \ref{3} is devoted to the application of QCD sum rules to the exclusive
decays of $B_c$ meson, where the estimates of form factors are given. The spin
symmetry in the $B_c$ decays is discussed and tested in Section \ref{4}. Our
numerical estimates of decay rates are presented in Sections \ref{5} and
\ref{6} for the semileptonic and non-leptonic modes, respectively. We summarize
our results in section \ref{7}.

\section{$\boldsymbol B_{\boldsymbol c} $ lifetime and inclusive decay
rates\label{2}}

The $B_c$-meson decay processes can be subdivided into three classes:

1) the $\bar b$-quark decay with the spectator $c$-quark,

2) the $c$-quark decay
with the spectator $\bar b$-quark and

3) the annihilation channel
$B_c^+\rightarrow l^+\nu_l (c\bar s, u\bar s)$, where $l=e,\; \mu,\; \tau$.

In the $\bar b \to \bar c c\bar s$ decays one separates also the
\underline{Pauli interference} with the $c$-quark from the initial state. In
accordance with the given classification, the total width is the sum over the
partial widths
$$
\Gamma (B_c\rightarrow X)=\Gamma (b\rightarrow X)
+\Gamma (c\rightarrow X)+\Gamma \mbox{(ann.)}+\Gamma\mbox{(PI)}.
$$
For the annihilation channel the  $\Gamma\mbox{(ann.)}$ width can be reliably
estimated in the framework of inclusive approach, where one takes the sum of
the leptonic and quark decay modes with account for the hard gluon corrections
to the effective four-quark interaction of weak currents. These corrections
result in the factor of $a_1=1.22\pm 0.04$. The width is expressed through the
leptonic constant of $f_{B_c}\approx 400$ MeV. This estimate of the
quark-contribution does not depend on a hadronization model, since a large
energy release of the order of the meson mass takes place. From the following
expression, one can see that the contribution by light leptons and quarks can
be neglected,
$$
\Gamma \mbox{(ann.)} =\sum_{i=\tau,c}\frac{G^2_F}{8\pi}
|V_{bc}|^2f^2_{B_c}M m^2_i (1-m^2_i/m^2_{Bc})^2\cdot C_i\;,
\label{d3}
$$
where $C_\tau = 1$ for the $\tau^+\nu_\tau$-channel and
$C_c =3|V_{cs}|^2a_1^2 $ for the $c\bar s$-channel.

As for the non-annihilation decays, in the approach of the {\sf Operator
Product Expansion} for the quark currents of weak decays \cite{OPEBc}, one
takes into account the $\alpha_s$-corrections to the free quark decays and uses
the quark-hadron duality for the final states. Then one considers the matrix
element for the transition operator over the bound meson state. The latter
allows one also to take into account the effects caused by the motion and
virtuality of decaying quark inside the meson because of the interaction with
the spectator. In this way the $\bar b\to \bar c c\bar s$ decay mode turns out
to be suppressed almost completely due to the Pauli interference with the charm
quark from the initial state. Besides, the $c$-quark decays with the spectator
$\bar b$-quark are essentially suppressed in comparison with the free quark
decays because of a large bound energy in the initial state.
\begin{table}[th]
\caption{The branching ratios of the $B_c$ decay modes calculated in
the framework of inclusive OPE approach, by summing up the exclusive modes in
the potential model \cite{PMK,PML} and according to the semi-inclusive
estimates in the sum rules of QCD and NRQCD \cite{KLO,KKL}. }
\label{t5}
\begin{center}
\begin{tabular}{|l|c|c|c|}
\hline
$B_c$ decay mode & OPE, \%  & PM, \% & SR, \%\\
\hline
$\bar b\to \bar c l^+\nu_l$ & $3.9\pm 1.0$  & $3.7\pm 0.9$  & $2.9\pm 0.3$\\
$\bar b\to \bar c u\bar d$  & $16.2\pm 4.1$ & $16.7\pm 4.2$ & $13.1\pm 1.3$\\
$\sum \bar b\to \bar c$     & $25.0\pm 6.2$ & $25.0\pm 6.2$ & $19.6\pm 1.9$\\
$c\to s l^+\nu_l$           & $8.5\pm 2.1$  & $10.1\pm 2.5$ & $9.0\pm 0.9$\\
$c\to s u\bar d$            & $47.3\pm 11.8$& $45.4\pm 11.4$& $54.0\pm 5.4$\\
$\sum c\to s$               & $64.3\pm 16.1$& $65.6\pm 16.4$& $72.0\pm 7.2$\\
$B_c^+\to \tau^+\nu_\tau$   & $2.9\pm 0.7$  & $2.0\pm 0.5$  & $1.8\pm 0.2$\\
$B_c^+\to c\bar s$          & $7.2\pm 1.8$  & $7.2\pm 1.8$  & $6.6\pm 0.7$\\
\hline
\end{tabular}
\end{center}
\label{inc}
\end{table}

In the framework of {\sf exclusive} approach, it is necessary to sum up widths
of different decay modes calculated in the potential models. While considering
the semileptonic decays due to the $\bar b \to \bar c l^+\nu_l$ and $c\to s
l^+\nu_l$ transitions, one finds that the hadronic final states are practically
saturated by the lightest bound $1S$-state in the $(\bar c c)$-system, i.e. by
the $\eta_c$ and $J/\psi$ particles, and the $1S$-states in the $(\bar b
s)$-system, i.e. $B_s$ and $B_s^*$, which can only enter the accessible
energetic gap.

Further, the $\bar b\to \bar c u\bar d$ channel, for example, can be calculated
through the given decay width of $\bar b \to \bar c l^+\nu_l$ with account
for the color factor and hard gluon corrections to the four-quark
interaction. It can be also obtained as a sum over the widths of decays
with the $(u\bar d)$-system bound states.

The results of calculation for the total $B_c$ width in the inclusive OPE and
exclusive PM approaches give the values consistent with each other, if one
takes into account the most significant uncertainty related to the choice
of quark masses (especially for the charm quark), so that finally, we have
\begin{equation}
\left.\tau[B_c^+]\right._{\mbox{\small\sc ope,\,pm}}= 0.55\pm 0.15\; \mbox{ps,}
\end{equation}
which agrees with the measured value of $B_c$ lifetime.

\begin{figure}[th]
\setlength{\unitlength}{0.5mm}
\begin{center}
\begin{picture}(130,95)
\put(0,5){\epsfxsize=120\unitlength \epsfbox{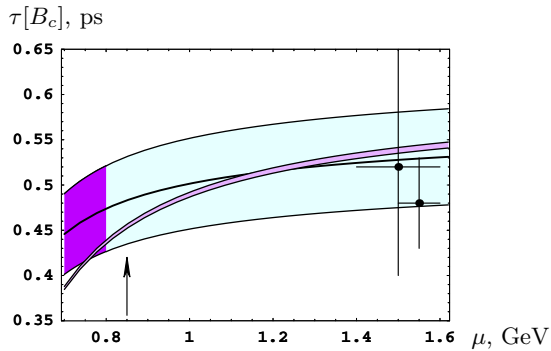}}
\put(0,92){$\tau[{B_c}] $, ps} \put(123,7){$\mu$, GeV}
\end{picture}
\end{center}
\caption{The $B_c$ lifetime calculated in QCD sum rules versus the scale of
hadronic weak lagrangian in the decays of charmed quark. The wide shaded region
shows the uncertainty of semi-inclusive estimates, the dark shaded region is
the preferable choice as given by the lifetimes of charmed mesons. The dots
represent the values in OPE approach taken from ref. \cite{OPEBc}. The narrow
shaded region represents the result obtained by summing up the exclusive
channels with the variation of hadronic scale in the decays of beauty
anti-quark in the range of $1 <\mu_b < 5$ GeV. The arrow points to the
preferable prescription of $\mu =0.85$ GeV as discussed in \cite{KKL}.}
\label{life}
\end{figure}

The OPE estimates of inclusive decay rates agree with recent semi-inclusive
calculations in the sum rules of QCD and NRQCD \cite{KLO,KKL}, where one
assumed the saturation of hadronic final states by the ground levels in the
$c\bar c$ and $\bar b s$ systems as well as the factorization allowing one to
relate the semileptonic and hadronic decay modes. The coulomb-like corrections
in the heavy quarkonia states play an essential role in the $B_c$ decays and
allow one to remove the disagreement between the estimates in sum rules and
OPE. In contrast to OPE, where the basic uncertainty is given by the variation
of heavy quark masses, these parameters are fixed by the two-point sum rules
for bottomonia and charmonia, so that the accuracy of SR calculations for the
total width of $B_c$ is determined by the choice of scale $\mu$ for the
hadronic weak lagrangian in decays of charmed quark. We show this dependence in
Fig. \ref{life}, where $\frac{m_c}{2} < \mu < m_c$ and the dark shaded region
corresponds to the scales preferred by data on the charmed meson lifetimes.

Supposing the preferable choice of scale in the $c\to s$ decays of $B_c$ to be
equal to $\mu^2_{B_c} \approx (0.85\; {\rm GeV})^2$, putting $a_1(\mu_{B_c})
=1.20$ and neglecting the contributions caused by nonzero $a_2$ in the charmed
quark decays \cite{KKL}, in the framework of semi-inclusive sum-rule
calculations we predict
\begin{equation}
\left.\tau[B_c]\right._{\mbox{\small\sc sr}} = 0.48\pm 0.05\;{\rm ps},
\end{equation}
which agrees with the direct summation of exclusive channels calculated in the
next sections. In Fig. \ref{life} we show the exclusive estimate of lifetime,
too.

\section{Machinery of QCD sum rules\label{3}}
In this section we present basic points for the procedure of calculating the
form factors for the various exclusive semileptonic decays of $B_c$. We omit
some technical details, which can be found in a number of references
appropriately cited.
\subsection{Form factors}
We accept the following convention on the normalization of wave functions for
the hadron states under study, i.e. for the pseudoscalar ($P$) and vector ($V$)
states:
\begin{eqnarray}
&& \langle 0|{\cal J}_\mu|P\rangle =  -{\rm i}\, f_P\,p_\mu,\label{pseudo}\\[1mm]
&& \langle 0|{\cal J}_\mu|V\rangle =  \epsilon_\mu\, f_V\,M_V,
\end{eqnarray}
where $f_{P,\,V}$ denote the leptonic constants, so that they are positive,
$$
f_{P,\,V} > 0,
$$
$p_\mu$ is a four-momentum of the hadron, $\epsilon_\mu$ is a polarization
vector of $V$, $M_V$ is its mass, and the current is composed of the valence
quark fields constituting the hadron
$$
{\cal J}_\mu = \bar q_1\, \gamma_\mu(1-\gamma_5)\,q_2.
$$
In this respect we can easily apply the ordinary Feynman rules for
the calculations of diagrams, so that the quark-meson vertices in
the decay channel are chosen with the following spin structures:
$$
\Gamma_P = \frac{\rm i}{\sqrt{2}}\,\gamma_5,\qquad \Gamma_V = -
\frac{1}{\sqrt{2}}\,\epsilon_\mu.
$$
Then, we get general expressions for the hadronic matrix elements of weak
currents in the exclusive decays of $P\to P^\prime$ and $P\to V$ with the
definitions of form factors given by the formulae
\begin{eqnarray}
\langle P^\prime(p_2)|{\cal J}_\mu |P(p_1)\rangle &=& f_{+} p_{\mu} +
f_{-}q_{\mu},\\
\frac{1}{\rm i}\langle V(p_2)|{\cal J}_\mu|P(p_1)\rangle &=&
{\rm i} F_V\epsilon_{\mu\nu\alpha\beta}\epsilon^{*\nu} p^{\alpha}q^{\beta}+
F_0^A\epsilon_{\mu}^{*} + \nonumber \\ &&
F_{+}^{A}(\epsilon^{*} p_1)p_{\mu} +
F_{-}^{A}(\epsilon^{*} p_1)q_{\mu},
\end{eqnarray}
where $q_{\mu} = (p_1 - p_2)_{\mu}$ and $p_{\mu} = (p_1 + p_2)_{\mu}$. The form
factors $f_{\pm}$ are dimensionless, while $F_V$ and $F^A_{\pm}$ has a
dimension of inverse energy, $F^A_0$ is of the energy dimension. In the case of
nonrelativistic description for both initial and final meson states we expect
that
$$
\begin{array}{rrr}
f_+ > 0, & f_- < 0, & F_V > 0, \\[2mm]
F^A_0 > 0, & F^A_+ < 0, & F^A_- > 0.
\end{array}
$$
It is important to note that for the pseudoscalar state the
hermitian conjugation in (\ref{pseudo}) does not lead to the
change of sign in the right hand side of equation because of the
prescription accepted, since the conjugation of imaginary unit
takes place with the change of sign for the momentum of meson (the
transition from the out-state to in-one). The same speculations
show that the spin structure of matrix element in the quark loop
order does not involve a functional dependence of form factors on
the transfer momentum squared except of $F^A_0$, so that we expect
that the simplest modelling in the form of the pole dependence can
be essentially broken for $F^A_0$, while the other form factors
are fitted by the pole model in a reasonable way.

Following the standard procedure for the evaluation of form factors in the
framework of QCD sum rules \cite{SR3pt}, in the $B_c$ decays we consider the
three-point functions
\begin{widetext}
\begin{eqnarray}
&& \Pi_{\mu}(p_1, p_2, q^2) = {\rm i}^2 \int {\rm d}x\,{\rm d}y\,e^{{\rm
i}(p_2\cdot x - p_1\cdot y)} \cdot 
\langle 0|T\{\bar q_2(x)\gamma_5 q_1(x)\, {\cal J}_\mu(0)\, \bar b(y)\gamma_5
c(y)\}|0 \rangle,\\
&& \Pi_{\mu\nu}^{{\cal J}}(p_1, p_2, q^2) = {\rm i}^2 \int {\rm d}x\,{\rm
d}y\,e^{{\rm i}(p_2\cdot x - p_1\cdot y)} \cdot 
\langle 0|T\{\bar q_2(x)\gamma_{\mu} q_1(x)\, {\cal J}_\mu(0)\,
\bar b(y)\gamma_5 c(y)\}|0\rangle,
\end{eqnarray}
\end{widetext}
where $\bar q_2(x)\gamma_5 q_1(x)$ and $\bar q_2(x)\gamma_{\nu}q_1(x)$ denote
the interpolating currents for the final states mesons.

The Lorentz structures in the correlators can be written down as
\begin{eqnarray}
\Pi_{\mu}~~ &=& \Pi_{+}(p_1 + p_2)_{\mu} + \Pi_{-}q_{\mu},\label{R1}\\
\Pi_{\mu\nu}^{\cal J} &=& {\rm
i}\,\Pi_V\epsilon_{\mu\nu\alpha\beta}p_2^{\alpha}p_1^{\beta}+
\Pi_{0}^{A}g_{\mu\nu} + \Pi_{1}^{A}p_{2, \mu}p_{1, \nu} +  \nonumber\\ &&
\Pi_{2}^{A}p_{1, \mu}p_{1, \nu} + \Pi_{3}^{A}p_{2, \mu}p_{2, \nu} +
\Pi_{4}^{A}p_{1, \mu}p_{2, \nu}.
\label{R2}
\end{eqnarray}
\noindent
The form factors $f_{\pm}$, $F_V$, $F_{0}^{A}$ and $F_{\pm}^{A}$ are determined
from the amplitudes $\Pi_{\pm}$, $\Pi_V$, $\Pi_{0}^{A}$ and $\Pi_{\pm}^{A} =
\frac{1}{2}(\Pi_{1}^{A}\pm \Pi_{2}^{A})$, respectively. In
(\ref{R1})--(\ref{R2}) the scalar amplitudes $\Pi$ are the functions of
kinematical invariants, i.e. $\Pi = \Pi(p_1^2, p_2^2, q^2)$.

\begin{figure}[th]
\setlength{\unitlength}{0.6mm}
\begin{center}
\begin{picture}(100,75)
\put(0, 0){\epsfxsize=110\unitlength \epsfbox{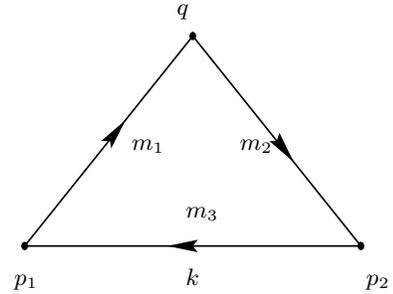}}
\put(85, 10){$p_{2}$}
\put(43,70){$q$}
\put(7, 10){$p_{1}$}
\put(45, 10){$k$}
\put(45, 25){$m_3$}
\put(33, 40){$m_1$}
\put(57, 40){$m_2$}
\end{picture}
\end{center}
\caption{The triangle diagram, giving the leading perturbative term in the OPE
expansion of three-point function.}
\label{Diag-fig}
\end{figure}

The leading QCD term is a triangle quark-loop diagram in Fig. \ref{Diag-fig},
for which we can write down the double dispersion representation at $q^2\leq
0$
$$
\begin{array}{ll}
\Pi_i^{\rm pert}(p_1^2, p_2^2, q^2) = & \displaystyle
-\frac{1}{(2\pi)^2}\int \frac{\rho_i^{\rm pert}(s_1, s_2, Q^2)}{(s_1 -
p_1^2)(s_2 - p_2^2)}{\rm d}s_1{\rm d}s_2 \\[2mm]
& +\mbox{~subtractions},
\end{array}
\label{pertdisp}
$$
where $Q^2 = -q^2 \geq 0$, and the integration region is determined by the
condition
\begin{equation}
-1 < \frac{2s_1s_2 + (s_1 + s_2 - q^2)(m_b^2 - m_c^2 - s_1)}
{\lambda^{1/2}(s_1, s_2, q^2)\lambda^{1/2}(m_c^2, s_1, m_b^2)} < 1,
\end{equation}
where $\lambda(x_1, x_2, x_3) = (x_1 + x_2 - x_3)^2 - 4x_1x_2$.
The expressions
for spectral densities $\rho_i^{\rm pert}(s_1, s_2, Q^2)$ were calculated in
\cite{KLO} and presented explicitly in Appendix A of ref. \cite{KKL}.

The connection to hadrons in the framework of QCD sum rules is obtained by
matching the resulting QCD expressions of current correlators with the spectral
representation, derived from a double dispersion relation at $q^2\leq 0$, so
that
$$
\begin{array}{ll}
\Pi_i(p_1^2, p_2^2, q^2) = & \displaystyle
-\frac{1}{(2\pi)^2}\int \frac{\rho_i^{\rm phys}(s_1, s_2, Q^2)}{(s_1 -
p_1^2)(s_2 - p_2^2)}{\rm d}s_1{\rm d}s_2 \\[2mm]
& +\mbox{~subtractions},
\end{array}
\label{physdisp}
$$
Assuming that the dispersion relation is well convergent, the physical spectral
functions are generally saturated by the ground hadronic states and a continuum
starting at some effective thresholds $s_1^{th}$ and $s_2^{th}$
$$
\rho_i^{\rm phys}(s_1, s_2, Q^2) = \rho_i^{\rm res}(s_1, s_2, Q^2) +
\rho_i^{\rm cont}(s_1, s_2, Q^2),
$$
where the resonance term is expressed through the product of leptonic constant
and form factor for the transition under consideration, so that
\begin{eqnarray}
\rho_i^{\rm res}(s_1, s_2, Q^2) &=& {\langle 0| \bar q_3 \left[{\gamma_{5}
\atop \gamma_{\mu}}\right] q_2|P^\prime(V) \rangle\, \langle B_c|\bar b\gamma_5
c|0)\rangle}\cdot \nonumber\\[2mm]
&&  F_i(Q^2)\,{(2\pi)^2 \delta(s_1-M_1^2) \delta(s_2-M_2^2)}
\nonumber\\[2mm]
&&+\mbox{higher~state~contributions}, \nonumber
\end{eqnarray}
where $M_{1,2}$ denote the masses of hadrons in the initial and final states.
The continuum of hadron states is usually approximated by the perturbative
absorptive part of $\Pi_i$ above the continuum threshold, that involves the
parametric dependence of sum rule results on the values of threshold energies.

Then, the expressions for the form factors $F_i$ can be derived by equating the
representations for the three-point functions $\Pi_i$ in terms of physical
densities and those of calculated in QCD, which means the formulation of sum
rules. In this work we use the scheme of moments, implying the calculation of
$n$-th derivatives of scalar correlators $\Pi(p_1^2,p_2^2,Q^2)$ over the
squared momenta of initial and final channels $p_1^2$ and $p_2^2$ at
$p_1^2=p_2^2=Q^2=0$.

\subsection{Coulomb corrections in the heavy quarkonia}
\begin{figure}[th]
\setlength{\unitlength}{0.6mm}
\begin{center}
\begin{picture}(110,110)
\put(0, 0){\epsfxsize=110\unitlength \epsfbox{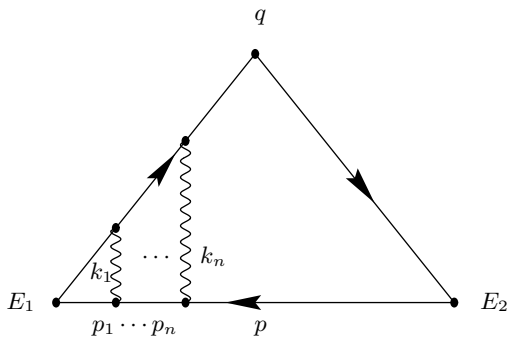}}
\put(0,40){$E_1$}
\put(105,40){$E_2$}
\put(55,35){$p$}
\put(55,104){$q$}
\put(19,35){$p_1\cdots p_n$}
\put(30,50){$\cdots$}
\put(18.5,46.5){$k_1$}
\put(43,50){$k_n$}
\end{picture}
\end{center}
\normalsize

\vspace*{-2cm}
\caption{The ladder diagram of the Coulomb-like interaction.}
\label{Coul-fig}
\end{figure}

For the heavy quarkonium $\bar b c$, where the relative velocity of quark
movement is small, an essential role is taken by the Coulomb-like
$\alpha_s/v$-corrections. They are caused by the ladder diagram, shown in
Fig. \ref{Coul-fig}. It is well known that an account for this corrections in
two-point sum rules numerically leads to a double-triple multiplication of Born
value of spectral density \cite{QCDSR}. In our case it leads to the finite
renormalization for $\rho_i$ \cite{KLO}, so that
\begin{equation}
 \rho^{c}_i={\cal C} \rho_i,
\label{ren}
\end{equation}
with
\begin{equation}
 {\cal C}^2=\left|\frac{\Psi^{\cal C}_{\bar b c}(0)}{\Psi^{\rm free}_{\bar b
 c}(0)}\right|^2=\frac{4\pi \alpha_s^{\cal
 C}}{3v}\,\frac{1}{\displaystyle 1-\exp\left(-\frac{4\pi\alpha_s^{\cal
 C}}{3v}\right)},
 \label{coul}
\end{equation}
where $v$ is the relative velocity of quarks in the $\bar b c$-system,
\begin{equation}
v=\sqrt{1-\frac{4 m_b m_c}{p_1^2-(m_b-m_c)^2}},
\end{equation}
and the coupling constant of effective coulomb interactions $\alpha_s^{\cal C}$
should be prescribed by the calculations of leptonic constants for the
appropriate heavy quarkonia as described in the next section.

A similar coulomb factor appears in the vertex of heavy quarks composing the
final heavy quarkonium, in the case of $\bar c c$.

\subsection{Leptonic constants of heavy quarkonia}
In order to fix such the parameters as the heavy quark masses and effective
couplings of coulomb exchange in the nonrelativistic systems of heavy quarkonia
with the same accuracy used in the three-point sum rules, we explore the
two-point sum rules of QCD for the systems of $\bar c c$, $\bar b c$ and $\bar
b b$. Thus, we take into account the quark loop contribution with the coulomb
factors like that of (\ref{coul}). We keep this procedure despite of current
status of NRQCD sum rules for the heavy quarkonium, wherein the three-loop
corrections to the correlators are available to the moment (see \cite{Hoang}
for review), since for the sake of consistency, the calculations should be
performed in the same order for both the three-point and two-point correlators.
This fact follows from the expression for the resonance term in the three-point
correlator, so that the form factor of transition involves the normalization by
the leptonic constants extracted from the two-point sum rules. This procedure
is taken since calculations of two-loop corrections to the three-point
correlators are not available to the moment, unfortunately.

\begin{figure}[th]
\setlength{\unitlength}{0.6mm}
\begin{center}
\begin{picture}(100, 80)
\put(0,0){\epsfxsize=100\unitlength\epsfbox{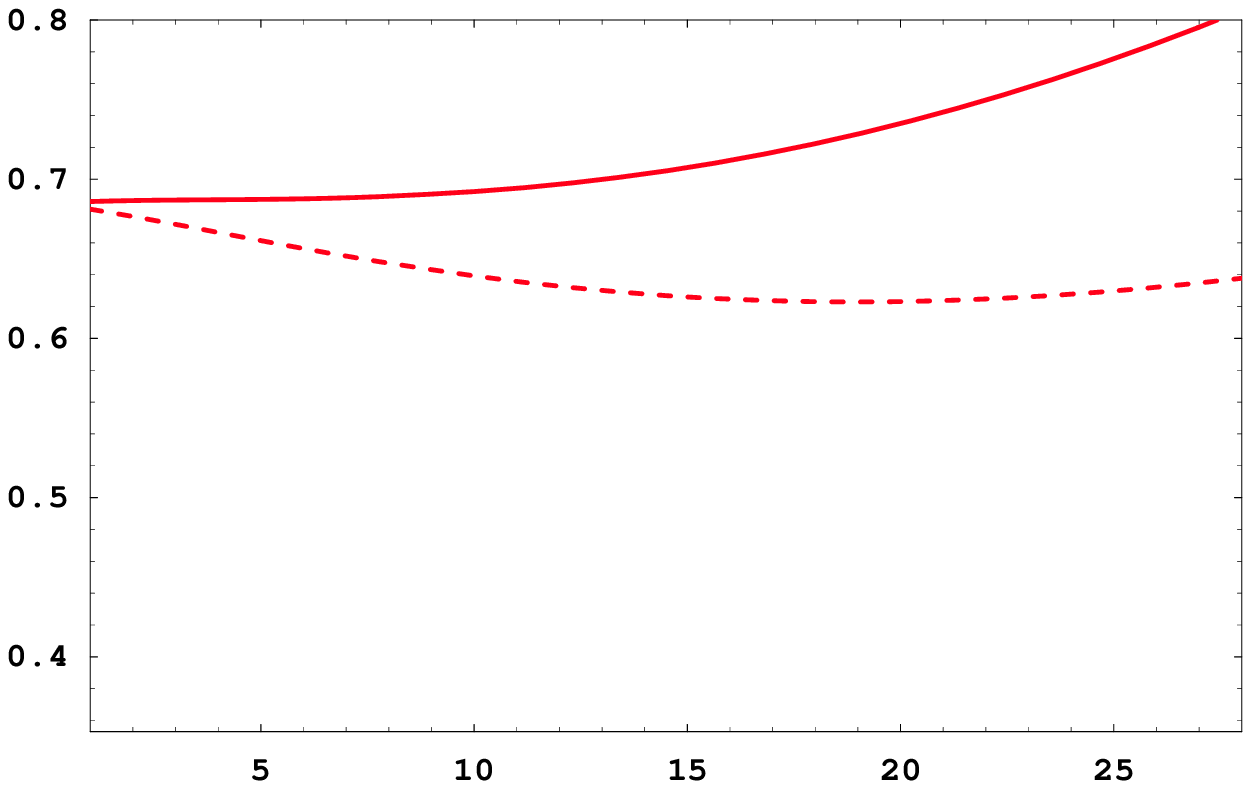}}
\put(0, 70){$f_{\Upsilon},~$MeV}
\put(100, 0){$n$}
\end{picture}
\end{center}
\caption{The two-point sum rules in the moment scheme for $\Upsilon$. The
dashed curve represents the result versus the moment number $n$ at
$m_b=4.63~$GeV, the solid line does at $m_b=4.59~$GeV.}
\label{Culfig}
\end{figure}

Then, the use of experimental values for the leptonic constants of charmonium
and bottomonium in addition to the consistent description of spectral function
moments in the two-point sum rules allows us to extract the effective couplings
of coulomb exchange as well as the heavy quark masses in the heavy quarkonium
channels. A characteristic picture for the description of leptonic constant for
the $1S$ vector state of bottomonium $\Upsilon$ is shown in Fig. \ref{Culfig}
as obtained in the scheme of moments for the spectral function.
In this calculations we take into account the contributions caused by the
higher excitations due to the known masses and ratios of leptonic constants to
that of ground state. These higher level terms are important at small $n$ in
order to get the stable description of leptonic constant for the ground state.
We have found that the normalization of leptonic constant is fixed by the
appropriate choice of effective constant for the coulomb exchange
$\alpha_s^{\cal C}$, while the stability is very sensitive to the prescribed
value of heavy quark mass.

The above procedure allows us to extract the following values of effective
constants:
$$
\alpha_s^{\cal C}[\bar b c]=0.45,\quad \alpha_s^{\cal C}[\bar c c]=0.60,\quad
\alpha_s^{\cal C}[\bar b b]=0.37,
$$
where we have used the experimental normalization of leptonic constant
\cite{PDG}
$$
f_{\psi}=(409\pm8)~{\rm MeV}, \qquad f_{\Upsilon}=(690\pm15)~{\rm  MeV.}
$$
and the scaling relation obtained in the quasilocal sum rules \cite{scale},
\begin{eqnarray}
 \frac{f_{\rm nS}^2}{M_{\rm nS}}\left(\frac{M_{\rm nS}(m_1+m_2)}{4 m_1
 m_2}\right)^2=\frac{c}{n},&&\label{scale}
 \end{eqnarray}
where $c$ is a constant value independent of the heavy flavours in the
quarkonium as well as its excitation number. Eq. (\ref{scale}) is valid for the
leptonic constants of ${\rm nS}$-levels in the heavy quarkonia composed of
heavy quarks with the masses $m_1$ and $m_2$. This relation yields also the
leptonic constant for the $B_c$ meson not yet measured to the moment, so that
the prediction\footnote{The consideration of leptonic constants for the heavy
quarkonia in the framework of potential approach was presented in
\cite{PMlept}.} reads off
$$
f_{B_c}=(400\pm15)~{\rm  MeV.}
$$
The heavy quark masses are given by
$$
m_b=4.60\pm0.02\,{\rm GeV}, \qquad
m_c=1.40\pm0.03\,{\rm GeV}.
$$
The physical meaning of above values for the heavy quark masses is determined
by the threshold posing the energy at which the coulomb spectrum starts, so
that it is very close to the so-called `potential subtracted masses' of heavy
quarks, $m^{\rm PS}$ known in the literature (see \cite{PS}) in the context of
renormalon in the perturbative pole mass \cite{renormalon}. We have found that
the numerical values obtained coincide with the appropriate $m^{\rm PS}$ within
the error bars.

\subsection{Leptonic constants of heavy-light mesons}
As was already explained the calculations in the three-point and two-point sum
rules should be performed in the same order of $\alpha_s$. So, although the
three-loop calculation are available for the leptonic constants of heavy-light
mesons to the moment \cite{fB}, we will use the quark-loop approximation (see
details in \cite{KKL}). In that case the renormalization constants of
quark-meson vertex in both the three-point and two-point sum rules cancel each
other in the expression for the transition form factors, while the absolute
normalization of leptonic constants of heavy-light mesons is irrelevant in this
respect, and it is defined by the loop corrections, which are rather
significant \cite{fB}.

Operationally, we put the following expression for the leptonic constant of
pseudoscalar meson containing the heavy quark $Q$
\begin{eqnarray}
f_{Q\bar q} &=& {\cal K}\cdot
E_c\,\sqrt{\frac{E_c}{m_Q}}\,\frac{\alpha_s^{1/9}(m_Q)}{\pi\sqrt{2}}\,
\left(\frac{m_Q}{M_{Q\bar q}}\right)^{3/2}\cdot \nonumber\\ &&
\left(1-\frac{2\alpha_s(m_Q)}{3\pi}+\frac{3}{88}\,\frac{E_c^2}{m_Q^2}-
\frac{\pi^2}{2}\,\frac{\langle \bar q q\rangle}{E_c^3}\right),
\end{eqnarray}
where we numerically accept
$$
\alpha_s(m_c) = 0.35,\qquad
\alpha_s(m_b) = 0.22,
$$
with the dimensional parameters
$$
E_c = 1.2\;\mbox{GeV},\qquad
\langle \bar u u\rangle = \langle \bar d d\rangle =
-(0.27)^3\;\mbox{GeV}^3,\qquad
$$
which was reasonably investigated in \cite{KKL}. We stress, first, that at the
fixed value of energy threshold $E_c$ in the heavy-light channel the total
dependence of transition form factors of $B_c$ into the heavy-light meson has a
local minimum, which ensures the stability of result with respect to a small
variation of $E_c$. Second, the $\cal K$-factor caused by the higher
loop-corrections is irrelevant for the extraction of form factors, although it
is large \cite{fB}.

Next, the fixed value of threshold energy $E_c$ determines the binding energy
of heavy quark in the meson, $\bar \Lambda\approx 0.63$ GeV, which yields the
same value of mass for the beauty quark, $m_b\approx 4.6$ GeV, as it was
determined from the analysis of two-point sum rules in the $\bar b b$ channel.
However, taking into account the second order corrections in $1/m_c$, we find
that the mass of charmed quark is shifted to the value of $m_c\approx 1.2$ GeV
in the heavy-light channel in comparison with the $\bar c c$ states. Thus, in
the transition of $B_c$ to the charmed meson we put the mass of spectator
charmed quark equal to $m_c\approx 1.2$ GeV, while for the transition into the
charmonium we accept $m_c\approx 1.4$ GeV, as discussed above.

Finally, we take the following ordinary ratios of leptonic constants for the
vector and pseudoscalar states and for the heavy-strange mesons:
$$
\frac{f_{B^*}}{f_B} \approx \frac{f_{D^*}}{f_D} \approx 1.11,\qquad
\frac{f_{B_s}}{f_B} \approx \frac{f_{D_s}}{f_D} \approx 1.16,
$$
which agree with the lattice computations \cite{latt}.

\subsection{Finite energy sum rules}
Putting the density of continuum hadronic states equal to that of calculated in
the perturbative QCD, we involve the dependence of calculations on the
threshold energy $s_{th}$ in the channels of two-quark currents, since this
procedure supposes the cancellation of integrals over the spectral functions in
both theoretical and physical parts of sum rules. At rather high numbers of
moments for the spectral density the contributions of excitations like $2S$,
are suppressed by the high powers of mass ratio $M_{1S}/M_{2S}$, so that the
local stability of form factor value on the variation of moment number allow us
extract the form factor for the transition of $1S\to 1S$, which is sufficient
for the $B_c$ decays in the heavy-light mesons. However, the estimates for the
decays into the quite narrow $2S$-resonances of charmonium cannot be obtained
in
this way.

We accept the concept of finite energy sum rules \cite{fesr} in order to
calculate the form factors for the decays of $B_c\to\bar c c[2S]$. In this
approach we suggest that with a rather good accuracy the contribution to the
correlator evaluated by the theoretical spectral density integrated out over
the finite energy interval containing the resonance under consideration gives
the approximation for the appropriate physical part excluding the contributions
of other resonances. In contrast to the global integration in the standard QCD
sum rules, where the result for the $1S$ channel reveals a weak dependence on
the continuum threshold, the choice of duality interval for the $2S$ level
involves a significant variation of the result on the level of 20\%. In order
to minimize the variation we explore the principle of stability, implying the
presence of local minimum in the ratios of form factors for the decays into the
$1S$ and $2S$ states of charmonium: $F_{B_c\to \bar c c[1S]}/F_{B_c\to \bar c
c[2S]}$.

\begin{figure}[th]
\setlength{\unitlength}{0.6mm}
\begin{center}
\begin{picture}(100,80)
\put(0,0){\epsfxsize=110\unitlength \epsfbox{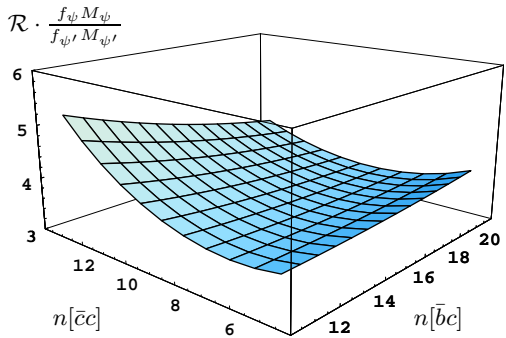}}
\put(10,5){$n[\bar c c]$} \put(90,5){$n[\bar b c]$}
\put(0,70){${\cal R}\cdot\frac{f_\psi M_{\psi}}{f_{\psi^\prime}
M_{\psi^\prime}}$}
\end{picture}
\end{center}
\caption{The ratio of form factors ${\cal R}$ for the transitions of $B_c$
meson into the vector charmonium states versus the numbers of moments in the
channels of $\bar b c$ and $\bar c c$.}
\label{2s}
\end{figure}

In detail, we parameterize the energy $E$ in the bound state channels by the
relative momentum $k$, so that
$$
E = \frac{k^2}{2 m_{\rm red}},
$$
where $m_{\rm red}$ is the reduced mass of quarks composing the meson. In the
QCD sum rules yielding the form factors for the $F_{B_c\to \bar c c[1S]}$
transition we put the initial and final values of momentum determining the
resonance range by
$$
k_i[1S]=0,\qquad k_{th}[1S]=1.2\;\mbox{GeV},
$$
while in the finite energy sum rules yielding the form factors for the
$F_{B_c\to \bar c c[2S]}$ transition we put
$$
k_i[2S]=1.09\;\mbox{GeV},\qquad k_{th}[2S]=1.2\;\mbox{GeV},
$$
which fits the region of $2S$ charmonium. At the above values of thresholds the
ratio of form factors
$$
{\cal R} = \frac{F_0^A(B_c\to \bar c c[1S])}{F_0^A(B_c\to \bar c
c[2S])}
$$
as well as both form factors themselves reveal the stability
versus the variation of moment numbers as shown in Fig. \ref{2s}.
The ratios for the other form factors  multiplied by the kinematic
factors depending on the total spin $j=0,1$ of recoil meson
$$
\kappa_{j} = \frac{f_{1S} M_{1S}^{2-j}}{f_{2S} M_{2S}^{2-j}}
$$
as evaluated in the framework of finite energy sum rules are
presented in Table \ref{rat}.

\begin{table}[th]
\caption{The ratio of form factors ${\cal R}_{2S}$ in the decays
of $B_c$ into the $1S$ and $2S$ charmonium states.}
\begin{tabular}{|c|c|c|c|c|c|c|}
\hline
Form factor & $f_+$ & $f_-$ & $F_V$ & $F_0^A$ & $F_+^A$ & $F_-^A$ \\
\hline ${\cal R}_{2S}\cdot\kappa_j$ & 3.9 & 2.2 & 3.1 & 3.5 & 4.9
& 2.3
\\
\hline
\end{tabular}
\label{rat}
\end{table}

Thus, we have found that the amplitudes squared for the decays of $B_c$ into
the $2S$ charmonium are suppressed by an order of magnitude in comparison with
the appropriate values for the decay into the ground pseudoscalar and vector
states.

\subsection{Numerical values}

\begin{figure}[th]
\setlength{\unitlength}{0.6mm}
\begin{center}
\begin{picture}(120,80)
\put(0,0){\epsfxsize=110\unitlength \epsfbox{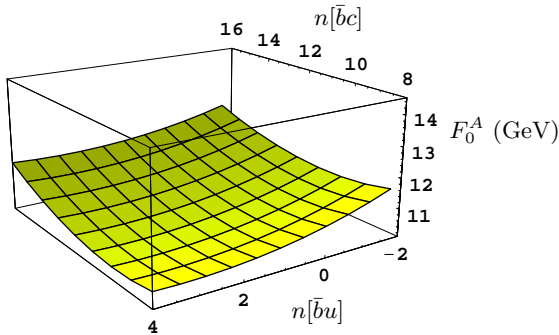}}
\put(75,70){$n[\bar b c]$} \put(70,5){$n[\bar b u]$}
\put(105,45){$F_0^A$ (GeV)}
\end{picture}
\end{center}
\caption{The form factor $F_0^A$ (GeV) at $q^2=0$ for the
transition of $B_c$ meson into the vector $B^*$ state versus the
numbers of moments in the channels of $\bar b c$ and $\bar b u$.}
\label{B}
\end{figure}

The characteristic dependence of sum rule results in the scheme of moments for
the spectral density is shown in Fig. \ref{B}, wherein we see the region of
stability versus the variation of moment numbers in both channels of initial
and final hadron states. Similar results are obtained for the other form
factors of decays under consideration.

\begin{table*}[th]
\caption{The form factors of various transitions calculated in the framework of
QCD sum rules at $q^2=0$ in comparison with the estimates in the potential
model (PM) of \cite{PMK}.}
\label{form}
\begin{tabular}{|c|c|c|c|c|c|c|}
\hline
Transition & $f_+$, [PM] & $f_-$, [PM] & $F_V$, [PM] (GeV$^{-1}$) & $F_0^A$,
[PM] (GeV) & $F_+^A$, [PM] (GeV$^{-1}$) & $F_-^A$, [PM] (GeV$^{-1}$) \\
\hline
$B_c\to B_s^{(*)}$ & 1.3, [1.1] & -5.8, [-5.9] & 1.1, [1.1] & 8.1, [8.2] &
0.2, [0.3] & 1.8, [1.4] \\
$B_c\to B^{(*)}$ & 1.27, [1.38] & -7.3, [-7.3] & 1.35, [1.37] & 9.8, [9.4] &
0.35, [0.36] & 2.5, [1.75] \\
$B_c\to D^{(*)}$ & 0.32, [0.29] & -0.34, [-0.37] & 0.20, [0.21] & 3.6, [3.6] &
-0.062, [-0.060] & 0.10, [0.16] \\
$B_c\to D_s^{(*)}$ & 0.45, [0.43] & -0.43, [-0.56] & 0.24, [0.27] & 4.7, [4.7]
&
-0.077, [-0.071] & 0.13, [0.20] \\
$B_c\to \bar c c[1S]$ & 0.66, [0.7]~ & -0.36, [-0.38] & 0.11, [0.10] & 5.9,
[6.2] & -0.074, [-0.070] & 0.12, [0.14] \\
\hline
\end{tabular}
\end{table*}
\begin{table*}[th]
\caption{The pole masses used in the model for the form factors in
various transitions.}
\label{pole}
\begin{tabular}{|l|c|c|c|c|c|c|}
\hline
Transition & $M_{\rm pole}[f_+]$, GeV & $M_{\rm pole}[f_-]$, GeV & $M_{\rm
pole}[F_V]$, GeV & $M_{\rm pole}[F_0^A]$, GeV & $M_{\rm pole}[F_+^A]$, GeV &
$M_{\rm pole}[F_-^A]$, GeV \\
\hline
$B_c\to \bar c c$ & 4.5 & 4.5 & 4.5 & 4.5 & 4.5 & 4.5 \\
$B_c\to B_s^{(*)}$ & 1.8 & 1.8 & 1.8 & 1.8 & 1.8 & 1.8 \\
$B_c\to B^{(*)}$ & 1.7 & 1.7 & 2.2 & 3.2 & 2.2 & 3.2 \\
$B_c\to D^{(*)}$ & 5.0 & 5.0 & 6.2 & $\infty$ & 6.2 & 6.2 \\
$B_c\to D_s^{(*)}$ & 5.0 & 5.0 & 6.2 & $\infty$ & 6.2 & 6.2 \\
\hline
\end{tabular}
\end{table*}
Our estimates are summarized in Table \ref{form}, where for the
sake of comparison we expose the results obtained in the potential
model \cite{PMK}, which parameters are listed in Appendix B of
ref.\cite{KKL}. In the potential model the most reliable results
are expected at zero recoil of meson in the final state of
transition, since the wave functions are rather accurately
calculable at small virtualities of quarks composing the meson. We
take the predictions of the potential model at zero recoil and
evolve the values of form factors to zero transfer squared in the
model with the pole dependence
$$
F_i(q^2) = \frac{F_i(0)}{1-q^2/{M_{i,\,{\rm pole}}^2}},
$$
making use of numerical values of $M_{i,\,{\rm pole}}$ shown in Table
\ref{pole}. We stress the fact that the potential model points to the
approximately constant value of the form factor $F_0^A$ because of additional
kinematical dependence in the transition of $B_c\to D^*$ and $B_c\to D_s^*$.

The sum rule estimates of form factors are taken at zero transfer squared,
while the dependence on $q^2$ is beyond the reliable accuracy of the method,
particularly, because at positive transfer squared some non-Landau
singularities could be important. So, the slopes of form factors are not under
control in the sum rules. Nevertheless, due to a suggestion on the pole
dependence we can restore the behaviour of form factors in the whole physical
region with the parameters shown in Table \ref{pole}. Of course, the numerical
values of pole masses can vary in reasonable ranges, which involves the
uncertainty in the estimates of width about 5--10\%, depending on the energy
release: at a small energy release the errors of modelling become less.

\section{Relations between the form factors\label{4}}

In the limit of infinitely heavy quark mass, the NRQCD and HQET lagrangians
possess the spin symmetry, since the heavy quark spin is decoupled in the
leading approximation. The most familiar implication of such the symmetry is
the common Isgur-Wise function determining the form factors in the semileptonic
decays of singly heavy hadrons.

In contrast to the weak decays with the light spectator quark, the
$B_c$ decays to both the charmonia $\psi$ and $\eta_c$ and
$B_s^{(*)}$ involve the heavy spectator, so that the spin symmetry
works only at the recoil momenta close to zero, where the
spectator enters the heavy hadron in the final state with no hard
gluon rescattering. Hence, in a strict consideration we expect the
relations between the form factors in the vicinity of zero recoil.
The normalization of common form factor is not fixed, as was in
decays of hadrons with a single heavy quark, since the heavy
quarkonia wave-functions are flavour-dependent. Nevertheless, in
practice, the ratios of form factors as fixed at a given zero
recoil point are broken only by the different dependence on the
transfer squared, that is not significant in real numerical
estimates in the restricted region of physical phase space.

As for the implications of spin symmetry for the form factors of decay, in the
soft limit for the transitions $B_c^+\to \psi(\eta_c) e^+\nu$
\begin{eqnarray}
v_1^{\mu} &\neq & v_2^{\mu}, \label{cs}\\
w &=& v_1\cdot v_2\to 1,\nonumber
\end{eqnarray}
where $v_{1,2}^{\mu} = p_{1,2}^{\mu}/\sqrt{p_{1,2}^2}$ are the four-velocities
of heavy quarkonia in the initial and final states, we derive the relations
\cite{KLO}
\begin{widetext}
\begin{eqnarray}
f_{+}(c_1^{P}\cdot{\cal M}_2 - c_2^{P}{\cal M}_1) - f_{-}(c_1^{P}\cdot{\cal
M}_2 + c_2^{P}\cdot{\cal M}_1) = 0, && \quad
F_{0}^{A}\cdot c_V - 2 c_{\epsilon}\cdot F_V{\cal M}_1{\cal M}_2  = 0,
\label{Fsym}\\[2mm]
F_{0}^{A}(c_1 + c_2) - c_{\epsilon}{\cal M}_1 (F_{+}^{A}({\cal M}_1 + {\cal
M}_2) +  F_{-}^{A}({\cal M}_1 - {\cal M}_2))  = 0, &&\quad
F_{0}^{A}c_1^{P} + c_{\epsilon}\cdot{\cal M}_1(f_{+} + f_{-})  = 0,
\end{eqnarray}
where
\begin{equation}
\begin{array}{lll}
c_{\epsilon} = -2,&~~
\displaystyle c_1 = -\frac{m_3(3m_1 + m_3)}{4m_1m_2},&~~
\displaystyle c_2 = \frac{1}{4m_1m_2}(4m_1m_2 + m_1m_3 + 2m_2m_3 +
m_3^2),\\[4mm]
\displaystyle c_1^{P} = 1 + \frac{m_3}{2m_1} - \frac{m_3}{2m_2},&~~
\displaystyle c_2^{P} = 1 - \frac{m_3}{2m_1} + \frac{m_3}{2m_2},&~~
\displaystyle c_V = -\frac{1}{2m_1m_2}(2m_1m_2 + m_1m_3 + m_2m_3),
\end{array}
\end{equation}
\end{widetext}
so that $m_1$ is the mass of decaying quark, $m_2$ is the quark mass of decay
product, and $m_3$ is the mass of spectator quark, while ${\cal M}_1=m_1+m_2$,
${\cal M}_2=m_2+m_3$.

The SR estimates of form factors show a good agreement with the relations,
whereas the deviations can be basically caused by the difference in the
$q^2$-evolution of form factors from the zero recoil point, that can be
neglected within the accuracy of SR method for the transitions of $B_c\to \bar
c c$ as shown in \cite{KLO}.

In the same limit for the semileptonic modes with a single heavy quark in the
final state we find that the ambiguity in the `light quark
propagator' 
(strictly, we deal with the uncertainty in the spin structure of amplitude
because of light degrees of freedom) 
restricts the number of relations, and we derive
\begin{widetext}
\begin{equation}
f_{+}(\bar c_1^{P}\cdot{\cal M}_2 - \bar c_2^{P}{\cal M}_1) - f_{-}(\bar
c_1^{P}\cdot{\cal M}_2 + \bar c_2^{P}\cdot{\cal M}_1) = 0,\quad
F_{0}^{A}\cdot \bar c_V - 2 \bar c_{\epsilon}\cdot F_V{\cal M}_1{\cal M}_2 =
0,\quad \label{Fsymq}
F_{0}^{A}\bar c_1^{P} + \bar c_{\epsilon}\cdot{\cal M}_1(f_{+} + f_{-}) = 0,
\end{equation}
where
\begin{equation}
\bar c_{\epsilon} = -2,\quad
\bar c_V = -1-\tilde B-\frac{m_3}{2m_1},\quad
\bar c_1^{P} = 1-\tilde B+\frac{m_3}{2m_1},\quad
\bar c_2^{P} = 1+\tilde B-\frac{m_3}{2m_1},
\end{equation}
\end{widetext}
so that $m_2$ is the mass of the light quark. The parameter $\tilde B$ has the
form
\begin{equation}
\tilde B=-\frac{2m_1+m_3}{2m_1}+\frac{4m_3(m_1+m_3)F_V}{F_0^A}.
\end{equation}
The $1/m_Q$-deviations from the symmetry relations in the decays of $B_c^+\to
B_s^{(*)} e^+\nu$ are about 10-15 \%, as found in the QCD sum rules
considered in \cite{KKL}.

Next, we investigate the validity of spin-symmetry relations in the $B_c$
decays to $B^{(*)}$, $D^{(*)}$ and $D_s^{(*)}$. The results of estimates for
the $f_{\pm}$ evaluated by the symmetry relations with the inputs given by the
form factors $F_V$ and $F_0^A$ extracted from the sum rules are presented in
Table \ref{symm} in comparison with the values calculated in the framework of
sum rules.

\begin{table*}[th]
\caption{The comparison of sum rule results for the form factors $f_{\pm}$ with
the values obtained by the spin symmetry with the inputs of $F_0^A$ and $F_V$
extracted from the QCD sum rules. The values of pole masses are also
explicitly shown.}
\label{symm}
\begin{tabular}{|l|c|c|c|c|c|}
\hline
Transition & Form factor & Sum Rules & Spin symmetry & $M_{\rm pole}[f_{\pm}]$,
GeV & $M_{\rm pole}[F_V]$, GeV \\
\hline
$B_c\to D^{(*)}$ & $f_+$ & 0.32 & 0.31 & 4.8 & 6.2 \\
$B_c\to D^{(*)}$ & $f_-$ & -0.34 & -0.36 & 4.8 & 6.2 \\
$B_c\to D_s^{(*)}$ & $f_+$ & 0.45 & 0.45 & 4.8 & 6.2 \\
$B_c\to D_s^{(*)}$ & $f_-$ & -0.43 & -0.51 & 4.8 & 6.2 \\
$B_c\to B^{(*)}$ & $f_+$ & 1.27 & 1.53 & 1.5 & 2.6 \\
$B_c\to B^{(*)}$ & $f_-$ & -7.3 & -7.1 & 1.5 & 2.6 \\
\hline
\end{tabular}
\end{table*}
We have found that the uncertainty in the estimates is basically determined by
the variation of pole masses in the $q^2$-dependencies of form factors, which
govern the evolution from the zero recoil point to the zero transfer squared.
So, the variation of $M_{\rm pole}[f_{\pm}]$ in the range of $4.8-5$ GeV for
the transitions of $B_c\to D^{(*)}$ and $B_c\to D_s^{(*)}$ results in the
30\%-uncertainty in the form factors presented in Table \ref{symm}.
Analogously, the variation of $M_{\rm pole}[f_{\pm}]$ in the range of $1.5-1.9$
GeV for the transition of $B_c\to B^{(*)}$ results in the uncertainty about
35\%.

Note, that the combinations of relations given above reproduce the only
equality \cite{Jenk}, which was found for each mode in the strict limit of
$v_1=v_2$.

\section{Semileptonic and leptonic modes\label{5}}
\subsection{Semileptonic decays}
The semileptonic decay rates are underestimated in the QCD SR approach of ref.
\cite{QCDSRBc}, because large coulomb-like corrections were not taken into
account. The recent analysis of SR in \cite{KT,KLO,KKL} decreased the
uncertainty, so that the estimates agree with the calculations in the potential
models.

The absolute values of semileptonic widths are presented in Table
\ref{tp81} in comparison with the estimates obtained in potential
models.

In practice, the most constructive information is given by the
$\psi$ mode, since this charmonium is clearly detected in
experiments due to the pure leptonic decays \cite{cdf}. In
addition to the investigation of various form factors and their
dependence on the transfer squared, we would like to stress that
the measurement of decay to the excited state of charmonium, i.e.
$\psi^\prime$, could answer the question on the reliability of QCD
predictions for the decays to the excited states. We see that to
the moment the finite energy sum rules predict the width of
$B_c^+\to \psi^\prime l^+ \nu$ decays in a reasonable agreement
with the potential models if one takes into account an uncertainty
about 50\%.

\begin{table}[h]
\caption{
Exclusive widths of semileptonic $B_c^+$ decays, $\Gamma$ in
$10^{-15}$ GeV, the symbol $\star$ marks the result of this work.
}
\begin{center}
\begin{tabular}{|l|r|r|r|r|r|r|}
\hline
~~~~~Mode & $\Gamma$ [$\star$] & $\Gamma$ \cite{vary} & $\Gamma$
\cite{chch} & $\Gamma$ \cite{ivanov} & $\Gamma$ \cite{ISGW2} & $\Gamma$
\cite{narod}\\
\hline
 $B_c^+ \rightarrow \eta_c e^+ \nu$
& 11 &11.1 &14.2  & 14 &10.4 &8.6\\
 $B_c^+ \rightarrow \eta_c \tau^+ \nu$
& 3.3 &  &  & 3.8 & &2.9\\
 $B_c^+ \rightarrow \eta_c^\prime e^+ \nu$
& 0.28 &  & 0.73  &  &0.74 & \\
 $B_c^+ \rightarrow \eta_c^\prime \tau^+ \nu$
& 0.024 &  &  & & & \\
 $B_c^+ \rightarrow J/\psi e^+ \nu $
& 28 &30.2 &34.4 & 33 & 16.5  &18\\
 $B_c^+ \rightarrow J/\psi \tau^+ \nu $
& 7.0 & &  & 8.4 & &5.0 \\
 $B_c^+ \rightarrow \psi^\prime e^+ \nu $
& 1.36 & & 1.45 & &3.1 & \\
 $B_c^+ \rightarrow \psi^\prime \tau^+ \nu $
& 0.12 & & &  & & \\
 $B_c^+ \rightarrow  D^0 e^+ \nu $
& 0.059 & 0.049 & 0.094 & 0.26 & 0.026 & \\
 $B_c^+ \rightarrow  D^0 \tau^+ \nu $
& 0.032 & &  &0.14 & & \\
 $B_c^+ \rightarrow  D^{*0} e^+ \nu  $
& 0.27 & 0.192 & 0.269 & 0.49 & 0.053 & \\
 $B_c^+ \rightarrow  D^{*0} \tau^+ \nu  $
& 0.12 & & & 0.27 & & \\
\hline\hline
 $B_c^+ \rightarrow  B^0_s e^+ \nu  $
& 59 &14.3 &26.6 &29 &13.8 &15 \\
 $B_c^+ \rightarrow B_s^{*0} e^+ \nu  $
& 65 &50.4 &44.0 &37 &16.9 &34\\
  $B_c^+ \rightarrow B^0 e^+ \nu  $
& 4.9 &1.14 &2.30 &2.1 & & \\
 $B_c^+ \rightarrow B^{*0} e^+ \nu  $
& 8.5 &3.53 &3.32 &2.3 & & \\
\hline
\end{tabular}
\label{tp81}
\end{center}
\end{table}

\subsection{Leptonic decays}

The dominant leptonic decay of $B_c$ is given by the $\tau \nu_\tau$ mode (see
Table \ref{inc}). However, it has a low experimental efficiency of detection
because of hadronic background in the $\tau$ decays or a missing energy.
Recently, in refs. \cite{radlep} the enhancement of muon and electron channels
in the radiative modes was studied. The additional photon allows one to remove
the helicity suppression for the leptonic decay of pseudoscalar particle, which
leads, say, to the double increase of muonic mode.

\section{Non-leptonic modes\label{6}}

In comparison with the inclusive non-leptonic widths, which can be
estimated in the framework of quark-hadron duality (see Table \ref{inc}), the
calculations of exclusive modes usually involves the approximation of
factorization \cite{fact}, which, as expected, can be quite accurate for the
$B_c$, since the quark-gluon sea is suppressed in the heavy quarkonium. Thus,
the important parameters are the factors $a_1$ and $a_2$ in the non-leptonic
weak lagrangian, which depend on the normalization point suitable for the $B_c$
decays.

\begin{table}[h!tb]
\caption{Exclusive non-leptonic decay widths of the $B_c$ meson, $\Gamma$ in
$10^{-15}$ GeV, the symbol $\star$ marks the result of this work. The $\bar
b$-quark decays with $c$-quark spectator. }
\begin{center}
\begin{tabular}{|c|l|r|r|r|r|}
\hline
Class & ~~~~~Mode & $\Gamma$ [$\star$] & $\Gamma$ \cite{vary} & $\Gamma$
\cite{chch} & $\Gamma$ \cite{narod}\\
\hline
& $B_c^+ \rightarrow \eta_c \pi^+$
&$1.8 \,a_1^2$ &$1.59 \,a_1^2$ &  $2.07  \,a_1^2$ &  $1.47  \,a_1^2$   \\
& $B_c^+ \rightarrow \eta_c \rho^+$
&$4.5 \,a_1^2$ &$3.74 \,a_1^2$ &  $5.48  \,a_1^2$  &  $3.35  \,a_1^2$ \\
& $B_c^+ \rightarrow J/\psi \pi^+$
&$1.43 \,a_1^2$ &$1.22 \,a_1^2$ &  $1.97  \,a_1^2$  &  $0.82  \,a_1^2$ \\
& $B_c^+ \rightarrow J/\psi \rho^+$
&$4.37 \,a_1^2$ &$3.48 \,a_1^2$ &  $5.95  \,a_1^2$  &  $2.32  \,a_1^2$ \\
I
& $B_c^+ \rightarrow \eta_c K^+ $
&$0.15 \,a_1^2$ &$0.119 \,a_1^2$&  $0.161 \,a_1^2$  &  $0.15  \,a_1^2$ \\
& $B_c^+ \rightarrow \eta_c K^{*+}$
&$0.22 \,a_1^2$ &$0.200 \,a_1^2$&  $0.286 \,a_1^2$  &  $0.24  \,a_1^2$ \\
& $B_c^+ \rightarrow J/\psi K^+$
&$0.12 \,a_1^2$ &$0.090 \,a_1^2$&  $0.152 \,a_1^2$  &  $0.079  \,a_1^2$ \\
& $B_c \rightarrow J/\psi K^{*+}$
&$0.25 \,a_1^2$ &$0.197 \,a_1^2$&  $0.324 \,a_1^2$  &  $0.18  \,a_1^2$ \\
\hline\hline
& $B_c^+ \rightarrow D^+
\overline D^{\hspace{1pt}\raisebox{-1pt}{$\scriptscriptstyle 0$}}$
&$1.9 \,a_2^2$ &$0.633 \,a_2^2 $ & $0.664  \,a_2^2 $ & $2.72  \,a_2^2 $  \\
& $B_c^+ \rightarrow D^+
\overline D^{\hspace{1pt}\raisebox{-1pt}{$\scriptscriptstyle *0$}}$
&$2.75 \,a_2^2 $ &$0.762 \,a_2^2 $ & $0.695  \,a_2^2 $ & $2.10  \,a_2^2 $  \\
& $B_c^+ \rightarrow  D^{\scriptscriptstyle *+}
\overline D^{\hspace{1pt}\raisebox{-1pt}{$\scriptscriptstyle 0$}}$
&$1.8 \,a_2^2 $ &$0.289 \,a_2^2 $ & $0.653  \,a_2^2 $ & $0.86  \,a_2^2 $  \\
& $B_c^+ \rightarrow  D^{\scriptscriptstyle *+}
\overline D^{\hspace{1pt}\raisebox{-1pt}{$\scriptscriptstyle *0$}}$
&$12.0 \,a_2^2 $ &$0.854 \,a_2^2 $ & $1.080  \,a_2^2 $ & $1.32  \,a_2^2 $  \\
II & $B_c^+ \rightarrow D_s^+ \overline
D^{\hspace{1pt}\raisebox{-1pt}{$\scriptscriptstyle 0$}}$
&$0.18 \,a_2^2 $ &$0.0415 \,a_2^2 $& $0.0340 \,a_2^2 $ & \\
& $B_c^+ \rightarrow D_s^+
\overline D^{\hspace{1pt}\raisebox{-1pt}{$\scriptscriptstyle *0$}}$
&$0.25 \,a_2^2 $ &$0.0495 \,a_2^2 $& $0.0354 \,a_2^2 $ & \\
& $B_c^+ \rightarrow  D_s^{\scriptscriptstyle *+} \overline
D^{\hspace{1pt}\raisebox{-1pt}{$\scriptscriptstyle 0$}}$
&$0.17 \,a_2^2 $ &$0.0201 \,a_2^2 $& $0.0334 \,a_2^2 $ & \\
& $B_c^+ \rightarrow  D_s^{\scriptscriptstyle *+}
\overline D^{\hspace{1pt}\raisebox{-1pt}{$\scriptscriptstyle *0$}}$
&$0.93 \,a_2^2 $ &$0.0597 \,a_2^2 $& $0.0564 \,a_2^2 $ & \\
\hline
\end{tabular}
\label{I,II}
\end{center}
\end{table}

\begin{table*}[ht]
\caption{Exclusive non-leptonic decay widths of the $B_c$ meson, $\Gamma$ in
$10^{-15}$ GeV, the symbol $\star$ marks the result of this work. The $\bar
b$-quark decays involving the Pauli interference with the $c$-quark spectator.
}
\begin{center}
\begin{tabular}{|c|l|c|c|c|c|}
\hline
Class & ~~~~~Mode & $\Gamma$ [$\star$] & $\Gamma$ \cite{vary} & $\Gamma$
\cite{chch} & $\Gamma$ \cite{narod} \\
\hline & $B_c^+ \rightarrow \eta_c D_s^+$ &$( 2.39\,a_1+
3.50\,a_2)^2$ &$( 2.16\,a_1
+2.57\,a_2)^2$&$(1.13\,a_1+1.98\,a_2)^2$ &$(2.58\,a_1+3.40\,a_2)^2$   \\
& $B_c^+ \rightarrow \eta_c D_s^{*+}$
&$( 2.16\,a_1+ 2.51\,a_2)^2$ &$( 2.03\,a_1+2.16\,a_2)^2$&$( 1.04\,a_1+
1.90\,a_2)^2$  &$(2.36\,a_1+1.99\,a_2)^2$ \\
& $B_c^+ \rightarrow J/\psi D_s^+$
&$( 1.92\,a_1+ 3.11\,a_2)^2$ &$(  1.62\,a_1+ 1.72\,a_2)^2$&$( 1.02\,a_1+
1.95\,a_2)^2$  &$(1.65\,a_1+2.92\,a_2)^2$ \\
& $B_c^+ \rightarrow J/\psi D_s^{*+}$ &$ 13.3\,a_1^2+42.2\, a_1
a_2+48.3\,a_2^2$ &$( 3.13\,a_1+ 3.67\,a_2)^2$&$
$  &$(3.31\,a_1+3.89\,a_2)^2$ \\
III& $B_c^+ \rightarrow \eta_c D^+$ &$( 0.50\,a_1+0.56\,a_2)^2$
&$(0.485\, a_1+ 0.528\,a_2)^2$&$(0.193\,a_1 + 0.440
\,a_2)^2$ &$(0.47\,a_1+0.73\,a_2)^2$ \\
& $B_c^+ \rightarrow \eta_c D^{*+}$
&$( 0.44\,a_1+0.59\,a_2)^2$ &$(  0.466\,a_1+ 0.452\,a_2)^2$&$( 0.181 \,a_1+
0.430 \,a_2)^2$ &$(0.37\,a_1+0.66\,a_2)^2$ \\
& $B_c^+ \rightarrow J/\psi D^+$
&$( 0.41\,a_1+0.53\,a_2)^2$ &$(  0.372\,a_1+ 0.338\,a_2)^2$&$( 0.177 \,a_1+
0.442 \,a_2)^2$ &$(0.30\,a_1+0.44\,a_2)^2$ \\
& $B_c^+ \rightarrow J/\psi D^{*+}$ &$ 0.52\,a_1^2+1.50\, a_1
a_2+1.68\,a_2^2$ &$(0.686\,a_1+ 0.732\,a_2)^2$  &$
$  &$(0.48\,a_1+0.80\,a_2)^2$ \\
\hline
\end{tabular}
\label{III}
\end{center}
\end{table*}

The QCD SR estimates for the non-leptonic decays of charmed quark in $B_c$ give
the widths represented in Tables \ref{I,II} and \ref{III}. The agreement of
results with the values predicted by the potential models is rather good for
the direct transitions with no permutation of colour lines, i.e. the class I
processes with the factor of $a_1$ in the non-leptonic amplitude determined by
the effective lagrangian. In contrast, the sum rule predictions are
significantly enhanced in comparison with the values calculated in the
potential models for the transitions with the colour permutation, i.e. for the
class II processes with the factor of $a_2$.

Further, for the transitions, wherein the interference is
significantly involved, the class III processes, we find that the
absolute values of different terms given by the squares of $a_1$
and $a_2$ calculated in the sum rules are in agreement with the
estimates of potential models. We stress that under fixing the
definitions of hadron state phases as described in section
\ref{3}.A, we have found that the Pauli interference has
determined the negative sign of two amplitudes with $a_1$ and
$a_2$, however, the relevant Fierz transformation has led to the
complete cancellation of the Pauli interference effect, and the
relative sign of two amplitude in the modes under consideration is
positive in agreement with the results of potential models listed
in Table \ref{III}. Taking into account the negative value of
$a_2$ with respect to $a_1$, we see that all of decays shown in
Table \ref{III} should be suppressed in comparison with the case
of the interference switched off. The characteristic values of
effects caused by the interference is presented in Table
\ref{Pauli}, where we put the widths in the form
$$
\Gamma = \Gamma_0 + \Delta \Gamma,\qquad \Gamma_0=x_1\, a_1^2+x_2\, a_2^2,\quad
\Delta\Gamma =z a_1\, a_2.
$$
Then, we conclude that the interference can be straightforwardly
tested in the listed decays, wherein its significance reaches
about 50\%.

\begin{table}[ht]
\caption{The effect of interference in the exclusive non-leptonic
decay widths of the $B_c$ meson with the $c$-quark as spectator at
$a_1^b =1.14$ and $a_2^b=-0.20$.}
\begin{center}
\begin{tabular}{|l|c|}
\hline
~~~~~Mode & $\Delta\Gamma/\Gamma_0$, \% \\
\hline
$B_c^+ \rightarrow \eta_c D_s^+$
& -48 \\
$B_c^+ \rightarrow \eta_c D_s^{*+}$
& -39 \\
$B_c^+ \rightarrow J/\psi D_s^+$
& -53 \\
$B_c^+ \rightarrow J/\psi D_s^{*+}$
& -50 \\
$B_c^+ \rightarrow \eta_c D^+$
& -38\\
$B_c^+ \rightarrow \eta_c D^{*+}$
& -45 \\
$B_c^+ \rightarrow J/\psi D^+$
& -43 \\
$B_c^+ \rightarrow J/\psi D^{*+}$
& -46 \\
\hline
\end{tabular}
\label{Pauli}
\end{center}
\end{table}

At large recoils as in $B_c^+\to \psi \pi^+(\rho^+)$, the
spectator picture of transition can be broken by the hard gluon
exchanges \cite{Gers}. The spin effects in such decays were
studied in \cite{Pakh}. However, we emphasize that the significant
rates of $B_c$ decays to the P- and D-wave charmonium states point
out that the corrections in the second order of the heavy-quark
velocity in the heavy quarkonia under study could be quite
essential and they can suppress the corresponding decay rates,
since the relative momentum of heavy quarks inside the quarkonium
if different from zero should enhance the virtuality of gluon
exchange, which suppresses the decay amplitudes.

\begin{table}[ht]
\caption{Exclusive non-leptonic decay widths of the $B_c$ meson, $\Gamma$ in
$10^{-15}$ GeV, the symbol $\star$ marks the result of this work. The
$c$-quark decays with $\bar b$-quark spectator. }
\begin{center}
\begin{tabular}{|c|l|c|c|c|c|}
\hline
Class & ~~~~~Mode & $\Gamma$ [$\star$] & $\Gamma$ \cite{vary} & $\Gamma$
\cite{chch} & $\Gamma$ \cite{narod} \\
\hline
& $B_c^+ \rightarrow B_s^0 \pi^+$
&$ 167\,a_1^2$ &$ 15.8\,a_1^2$&$ 58.4\,a_1^2$ &$ 34.8\,a_1^2$   \\
& $B_c^+ \rightarrow B_s^0 \rho^+$
&$ 72.5\,a_1^2$ &$ 39.2\,a_1^2$&$ 44.8\,a_1^2$ &$ 23.6\,a_1^2$   \\
& $B_c^+ \rightarrow B_s^{*0} \pi^+$
&$ 66.3\,a_1^2$ &$ 12.5\,a_1^2$&$ 51.6\,a_1^2$ &$ 19.8\,a_1^2$   \\
& $B_c^+ \rightarrow B_s^{*0} \rho^+$
&$ 204\,a_1^2$ &$ 171.\,a_1^2$&$ 150.\,a_1^2$  &$ 123\,a_1^2$  \\
I & $B_c^+ \rightarrow B_s^0 K^+$
&$ 10.7\,a_1^2$ &$ 1.70\,a_1^2$&$ 4.20\,a_1^2$ &  \\
& $B_c^+ \rightarrow B_s^{*0} K^+$
&$ 3.8\,a_1^2$ &$ 1.34\,a_1^2$&$ 2.96\,a_1^2$  & \\
& $B_c^+ \rightarrow B_s^0 K^{*+}$
& &$ 1.06\,a_1^2$&$          $  & \\
& $B_c^+ \rightarrow B_s^{*0} K^{*+}$
& &$ 11.6\,a_1^2$&$          $  & \\
& $B_c^+ \rightarrow B^0 \pi^+$
&$ 10.6\,a_1^2$ &$ 1.03\,a_1^2$&$ 3.30\,a_1^2$ &$ 1.50\,a_1^2$   \\
& $B_c^+ \rightarrow B^0 \rho^+$
&$ 9.7\,a_1^2$ &$ 2.81\,a_1^2$&$ 5.97\,a_1^2$  &$ 1.93\,a_1^2$  \\
& $B_c^+ \rightarrow B^{*0} \pi^+$
&$ 9.5\,a_1^2$ &$ 0.77\,a_1^2$&$ 2.90\,a_1^2$  &$ 0.78\,a_1^2$  \\
& $B_c^+ \rightarrow B^{*0} \rho^+$
&$ 26.1\,a_1^2$ &$ 9.01\,a_1^2$&$ 11.9\,a_1^2$ &$ 6.78\,a_1^2$   \\
& $B_c^+ \rightarrow B^0 K^+$
&$ 0.70\,a_1^2$ &$ 0.105\,a_1^2$&$ 0.255\,a_1^2$ & \\
& $B_c^+ \rightarrow B^0 K^{*+}$
&$ 0.15\,a_1^2$ &$ 0.125\,a_1^2$&$ 0.180\,a_1^2$ & \\
& $B_c^+ \rightarrow B^{*0} K^+$
&$ 0.56\,a_1^2$ &$ 0.064\,a_1^2$&$ 0.195\,a_1^2$ & \\
& $B_c^+ \rightarrow B^{*0} K^{*+}$
&$ 0.59\,a_1^2$ &$ 0.665\,a_1^2$&$ 0.374\,a_1^2$ &
\\
\hline\hline
& $B_c^+ \rightarrow B^+ \overline{K^0}$
&$ 286\,a_2^2$ &$ 39.1\,a_2^2$&$ 96.5\,a_2^2$ &$ 24.0\,a_2^2$   \\
& $B_c^+ \rightarrow B^+ \overline{K^{*0}}$
&$ 64\,a_2^2$ &$ 46.8\,a_2^2$&$ 68.2\,a_2^2$  &$ 13.8\,a_2^2$  \\
& $B_c^+ \rightarrow B^{*+} \overline{K^0}$
&$ 231\,a_2^2$ &$ 24.0\,a_2^2$&$ 73.3\,a_2^2$  &$ 8.9\,a_2^2$  \\
& $B_c^+ \rightarrow B^{*+} \overline{K^{*0}}$
&$ 242\,a_2^2$ &$ 247\,a_2^2$&$ 141\,a_2^2$  &$ 82.3\,a_2^2$  \\
II & $B_c^+ \rightarrow B^+ \pi^0$
&$ 5.3\,a_2^2$ &$ 0.51\,a_2^2$&$ 1.65\,a_2^2$ &$ 1.03\,a_2^2$   \\
& $B_c^+ \rightarrow B^+ \rho^0$
&$ 4.4\,a_2^2$ &$ 1.40\,a_2^2$&$ 2.98\,a_2^2$  &$ 1.28\,a_2^2$  \\
& $B_c^+ \rightarrow B^{*+} \pi^0$
&$ 4.8\,a_2^2$ &$ 0.38\,a_2^2$&$ 1.45\,a_2^2$ &$ 0.53\,a_2^2$   \\
& $B_c^+ \rightarrow B^{*+} \rho^0$
&$ 13.1\,a_2^2$ &$ 4.50\,a_2^2$&$ 5.96\,a_2^2$  &$ 4.56\,a_2^2$  \\
\hline
\end{tabular}
\label{bI,II}
\end{center}
\end{table}

The widths of non-leptonic $c$-quark decays are listed in Table \ref{bI,II} in
comparison with the predictions of potential models. We see that the sum rule
estimates are greater than those of potential models. In this respect we check
that our calculations are consistent with the inclusive ones. So, we sum up
the calculated exclusive widths and estimate the total width of $B_c$ meson as
shown in Fig. \ref{life}, which points to a good agreement of our calculations
with those of OPE and semi-inclusive estimates.

Another interesting point is the possibility to extract the factorization
parameters $a_1$ and $a_2$ in the $c$-quark decays by measuring the branching
ratios
\begin{eqnarray}
\frac{\Gamma[B_c^+\to B^+\bar K^0]}{\Gamma[B_c^+\to B^0 K^+]} &=&
\frac{\Gamma[B_c^+\to B^+\bar K^{*0}]}{\Gamma[B_c^+\to B^0 K^{*+}]} = \nonumber
\\[3mm]
\frac{\Gamma[B_c^+\to B^{*+}\bar K^0]}{\Gamma[B_c^+\to B^{*0} K^+]} &=&
\frac{\Gamma[B_c^+\to B^{*+}\bar K^{*0}]}{\Gamma[B_c^+\to B^{*0} K^{*+}]} =
\label{fact}\\[3mm]
\frac{\Gamma_0}{\Gamma_+}&=&
\left|\frac{V_{cs}}{V_{cd}^2}\right|^2\,\left(\frac{a_2}{a_1}\right)^2.
\nonumber
\end{eqnarray}
This procedure can give the test for the factorization approach itself.

The suppressed decays caused by the flavor changing neutral currents were
studied in \cite{rare}.

The {\sf CP}-violation in the $B_c$ decays can be investigated in
the same manner as made in the $B$ decays. The expected {\sf
CP}-asymmetry of ${\cal A}(B_c^\pm \to J/\psi D^\pm)$ is about
$4\cdot 10^{-3}$, when the corresponding branching ratio is
suppressed as $10^{-4}$ \cite{CPv}. Thus, the direct study of {\sf
CP}-violation in the $B_c$ decays is practically difficult because
of low relative yield of $B_c$ with respect to ordinary $B$
mesons: $\sigma(B_c)/\sigma(B) \sim 10^{-3}$. A model-independent
way to extract the CKM angle $\gamma$ based on the measurement of
two reference triangles was independently offered by Masetti,
Fleischer and Wyler in \cite{CPv} by investigating the modes with
the neutral charmed meson in the final state.

Another possibility is the lepton tagging of $B_s$ in the $B_c^\pm\to B_s^{(*)}
l^\pm \nu$ decays for the study of mixing and {\sf CP}-violation in the $B_s$
sector \cite{Quigg}.

\begin{table*}[th]
\caption{Branching ratios of exclusive $B_c^+$ decays at the fixed choice of
factors: $a_1^c =1.20$ and $a_2^c=-0.317$ in the non-leptonic decays of $c$
quark, and $a_1^b =1.14$ and $a_2^b=-0.20$ in the non-leptonic decays of $\bar
b$ quark. The lifetime of $B_c$ is appropriately normalized by $\tau[B_c]
\approx 0.45$ ps.}
\label{common}
\begin{center}
\begin{tabular}{|l|r|}
\hline
~~~~~Mode & BR, \%\\
\hline
 $B_c^+ \rightarrow \eta_c e^+ \nu$
 & 0.75\\
 $B_c^+ \rightarrow \eta_c \tau^+ \nu$
 & 0.23\\
 $B_c^+ \rightarrow \eta_c^\prime e^+ \nu$
 & 0.020\\
 $B_c^+ \rightarrow \eta_c^\prime \tau^+ \nu$
 & 0.0016\\
 $B_c^+ \rightarrow J/\psi e^+ \nu $
 & 1.9\\
 $B_c^+ \rightarrow J/\psi \tau^+ \nu $
 & 0.48\\
 $B_c^+ \rightarrow \psi^\prime e^+ \nu $
 & 0.094 \\
 $B_c^+ \rightarrow \psi^\prime \tau^+ \nu $
 & 0.008\\
 $B_c^+ \rightarrow  D^0 e^+ \nu $
 & 0.004 \\
 $B_c^+ \rightarrow  D^0 \tau^+ \nu $
 & 0.002 \\
 $B_c^+ \rightarrow  D^{*0} e^+ \nu  $
 & 0.018  \\
 $B_c^+ \rightarrow  D^{*0} \tau^+ \nu  $
 & 0.008 \\
 $B_c^+ \rightarrow  B^0_s e^+ \nu  $
 & 4.03  \\
 $B_c^+ \rightarrow B_s^{*0} e^+ \nu  $
 & 5.06 \\
  $B_c^+ \rightarrow B^0 e^+ \nu  $
 & 0.34\\
 $B_c^+ \rightarrow B^{*0} e^+ \nu  $
 & 0.58 \\
 $B_c^+ \rightarrow \eta_c \pi^+$
 & 0.20\\
 $B_c^+ \rightarrow \eta_c \rho^+$
 & 0.42\\
 $B_c^+ \rightarrow J/\psi \pi^+$
 & 0.13\\
 $B_c^+ \rightarrow J/\psi \rho^+$
 & 0.40\\
 $B_c^+ \rightarrow \eta_c K^+ $
 & 0.013\\
 $B_c^+ \rightarrow \eta_c K^{*+}$
 & 0.020\\
\hline
\end{tabular}
\begin{tabular}{|l|r|}
\hline
~~~~~Mode & BR, \%\\
\hline
 $B_c^+ \rightarrow J/\psi K^+$
 & 0.011\\
 $B_c \rightarrow J/\psi K^{*+}$
 & 0.022\\
 $B_c^+ \rightarrow D^+
\overline D^{\hspace{1pt}\raisebox{-1pt}{$\scriptscriptstyle 0$}}$
 & 0.0053\\
 $B_c^+ \rightarrow D^+
\overline D^{\hspace{1pt}\raisebox{-1pt}{$\scriptscriptstyle *0$}}$
 & 0.0075\\
 $B_c^+ \rightarrow  D^{\scriptscriptstyle *+}
\overline D^{\hspace{1pt}\raisebox{-1pt}{$\scriptscriptstyle 0$}}$
 & 0.0049\\
 $B_c^+ \rightarrow  D^{\scriptscriptstyle *+}
\overline D^{\hspace{1pt}\raisebox{-1pt}{$\scriptscriptstyle *0$}}$
 & 0.033\\
 $B_c^+ \rightarrow D_s^+ \overline
D^{\hspace{1pt}\raisebox{-1pt}{$\scriptscriptstyle 0$}}$
 & 0.00048\\
 $B_c^+ \rightarrow D_s^+
\overline D^{\hspace{1pt}\raisebox{-1pt}{$\scriptscriptstyle *0$}}$
 & 0.00071\\
 $B_c^+ \rightarrow  D_s^{\scriptscriptstyle *+} \overline
D^{\hspace{1pt}\raisebox{-1pt}{$\scriptscriptstyle 0$}}$
 & 0.00045\\
 $B_c^+ \rightarrow  D_s^{\scriptscriptstyle *+}
\overline D^{\hspace{1pt}\raisebox{-1pt}{$\scriptscriptstyle *0$}}$
 & 0.0026\\
 $B_c^+ \rightarrow \eta_c D_s^+$
 & 0.28\\
 $B_c^+ \rightarrow \eta_c D_s^{*+}$
 & 0.27\\
 $B_c^+ \rightarrow J/\psi D_s^+$
 & 0.17\\
 $B_c^+ \rightarrow J/\psi D_s^{*+}$
 & 0.67\\
 $B_c^+ \rightarrow \eta_c D^+$
 & 0.015\\
 $B_c^+ \rightarrow \eta_c D^{*+}$
 & 0.010\\
 $B_c^+ \rightarrow J/\psi D^+$
 & 0.009\\
 $B_c^+ \rightarrow J/\psi D^{*+}$
 & 0.028\\
 $B_c^+ \rightarrow B_s^0 \pi^+$
 & 16.4\\
 $B_c^+ \rightarrow B_s^0 \rho^+$
 & 7.2\\
 $B_c^+ \rightarrow B_s^{*0} \pi^+$
 & 6.5\\
 $B_c^+ \rightarrow B_s^{*0} \rho^+$
 & 20.2\\
\hline
\end{tabular}
\begin{tabular}{|l|r|}
\hline
~~~~~Mode & BR, \%\\
\hline
 $B_c^+ \rightarrow B_s^0 K^+$
 & 1.06\\
 $B_c^+ \rightarrow B_s^{*0} K^+$
 & 0.37\\
 $B_c^+ \rightarrow B_s^0 K^{*+}$
 & --\\
 $B_c^+ \rightarrow B_s^{*0} K^{*+}$
 & --\\
 $B_c^+ \rightarrow B^0 \pi^+$
 & 1.06\\
 $B_c^+ \rightarrow B^0 \rho^+$
 & 0.96\\
 $B_c^+ \rightarrow B^{*0} \pi^+$
 & 0.95\\
 $B_c^+ \rightarrow B^{*0} \rho^+$
 & 2.57\\
 $B_c^+ \rightarrow B^0 K^+$
 & 0.07\\
 $B_c^+ \rightarrow B^0 K^{*+}$
 & 0.015\\
 $B_c^+ \rightarrow B^{*0} K^+$
 & 0.055\\
 $B_c^+ \rightarrow B^{*0} K^{*+}$
 & 0.058\\
 $B_c^+ \rightarrow B^+ \overline{K^0}$
 & 1.98\\
 $B_c^+ \rightarrow B^+ \overline{K^{*0}}$
 & 0.43\\
 $B_c^+ \rightarrow B^{*+} \overline{K^0}$
 & 1.60\\
 $B_c^+ \rightarrow B^{*+} \overline{K^{*0}}$
 & 1.67\\
 $B_c^+ \rightarrow B^+ \pi^0$
 & 0.037\\
 $B_c^+ \rightarrow B^+ \rho^0$
 & 0.034\\
 $B_c^+ \rightarrow B^{*+} \pi^0$
 & 0.033\\
 $B_c^+ \rightarrow B^{*+} \rho^0$
 & 0.09\\
 $B_c^+ \rightarrow \tau^+ \nu_\tau$
 & 1.6\\
 $B_c^+ \rightarrow c \bar s$
 & 4.9\\
\hline
\end{tabular}
\end{center}
\end{table*}

\section{Discussion and Conclusions\label{7}}

We have calculated the exclusive decay widths of the $B_c$ meson in the
framework of QCD sum rules and reviewed the current status of theoretical
predictions for the decays of $B_c$ meson. We have found that the various
approaches: OPE, Potential models and QCD sum rules, result in the close
estimates, while the SR as explored for the various heavy quark systems, lead
to a smaller uncertainty due to quite an accurate knowledge of the heavy quark
masses. So, summarizing we expect that the dominant contribution to the $B_c$
lifetime is given by the charmed quark decays ($\sim 70\%$), while the
$b$-quark decays and the weak annihilation add about 20\% and 10\%,
respectively. The coulomb-like $\alpha_s/v$-corrections play an essential role
in the determination of exclusive form factors in the QCD SR. The form factors
obey the relations dictated by the spin symmetry of NRQCD and HQET with quite
a good accuracy expected.

The predictions of QCD sum rules for the exclusive decays of $B_c$ are
summarized in Table \ref{common} at the fixed values of factors $a_{1,2}$ and
lifetime. In addition to the decay channels with the heavy charmonium $J/\psi$
well detectable through its leptonic mode, one could expect a significant
information on the dynamics of $B_c$ decays from the channels with a single
heavy mesons, if an experimental efficiency allows one to extract a signal from
the cascade decays. An interesting opportunity is presented by the relations
for the ratios in (\ref{fact}), which can shed light to characteristics of the
non-leptonic decays in the explicit form.

We have found that the $\bar b$ decay to the doubly charmed states gives
$$
\mbox{Br}[B_c^+\to \bar c c\,c\bar s] \approx 1.39\%,
$$
so that in the absolute value of width it can be compared with the estimate of
spectator decay \cite{OPEBc},
\begin{eqnarray}
\left.\Gamma[B_c^+\to \bar c c\,c\bar s]\right|_{\mbox{\sc sr}}
&\approx & 20\cdot
10^{-15}\,\mbox{GeV},\nonumber \\[2mm]
\left.\Gamma[B_c^+\to \bar c c\,c\bar s]\right|_{\rm spect.} &\approx & 90\cdot
10^{-15}\,\mbox{GeV}, \nonumber
\end{eqnarray}
and we find the suppression factor of about $1/4.5$. This result
is in agreement with the estimate in OPE \cite{OPEBc}, where a
strong dependence of negative term caused by the Pauli
interference on the normalization scale of non-leptonic weak
lagrangian was emphasized, so that at moderate scales one gets
approximately the same suppression factor, too.

To the moment we certainly state that the accurate direct
measurement of $B_c$ lifetime can provide us with the information
on both the masses of charmed and beauty quarks and the
normalization point of non-leptonic weak lagrangian in the $B_c$
decays (the $a_1$ and $a_2$ factors). The experimental study of
semileptonic decays and the extraction of ratios for the form
factors can test the spin symmetry derived in the NRQCD and HQET
approaches and decrease the theoretical uncertainties in the
corresponding theoretical evaluation of quark parameters as well
as the hadronic matrix elements, determined by the nonperturbative
effects caused by the quark confinement. The measurement of
branching fractions for the semileptonic and non-leptonic modes
and their ratios can inform on the values of factorization
parameters, which depend again on the normalization of
non-leptonic weak lagrangian. The charmed quark counting in the
$B_c$ decays is related to the overall contribution of $b$ quark
decays as well as with the suppression of $\bar b\to c\bar c \bar
s$ transition because of the destructive interference, which value
depends on the nonperturbative parameters (roughly estimated, the
leptonic constant) and non-leptonic weak lagrangian.

Thus, the progress in measuring the $B_c$ lifetime and decays could enforce
the theoretical understanding of what really happens in the heavy quark decays
at all.

We point also to the papers, wherein some aspects of $B_c$ decays and
spectroscopy were studied:
non-leptonic decays in \cite{verma}, polarization effects in the radiative
leptonic decays \cite{giri}, relativistic effects \cite{Nobes}, spectroscopy in
the systematic approach of potential nonrelativistic QCD in \cite{Brambilla},
nonperturbative effects in the semileptonic decays \cite{Mannel}, exclusive and
inclusive decays of $B_c$ states into the lepton pair and hadrons \cite{Ma},
rare decays in \cite{Ivanov2}, the spectroscopy and radiative decays in
\cite{Ebert}.

This work is supported in part by the Russian Foundation for Basic Research,
grants 01-02-99315, 01-02-16585, and 00-15-96645.

\hbadness=10000

\end{document}